\documentclass[preprint,showpacs,preprintnumbers,amsmath,amssymb,nofootinbib]{revtex4}

\usepackage{graphicx}
\usepackage{dcolumn}
\usepackage{bm}
\usepackage{subfigure}

\def\beq{\begin{equation}}
\def\eeq{\end{equation}}
\def\bea{\begin{eqnarray}}
\def\eea{\end{eqnarray}}

\def\met{\ensuremath{\not\!\!{E_{T}}}}

\newcommand{\lsim}{ \mathop{}_{\textstyle \sim}^{\textstyle <} }

\usepackage{color}

\begin{document}

\preprint{UCI-HEP-TR-2011-22}

\title{LHC Bounds on UV-Complete Models of Dark Matter}

\author{Jessica Goodman}
\author{William Shepherd}

\affiliation{Department of Physics \& Astronomy, University of California, Irvine, CA 92697}

\date{\today}

\begin{abstract}
We analyze the sensitivity of searches for dark matter in the jets and missing
energy channel in the case where the particle mediating interactions between
hadronic matter and DM is collider accessible.
We consider all tree level UV completions of interactions between fermion DM 
and quarks which contribute to direct detection, and derive
bounds which apply to elastic or inelastic scattering dark matter explanations
of direct detection signals.  We find that studies based on effective operators
give robust bounds when the mediator is heavy enough to resonantly produce 
the final state in question.
\end{abstract}


\maketitle

\section{Introduction}

The need for extra non-baryonic mass to explain galactic and supergalactic dynamics is well
known from astrophysical observations \cite{Komatsu:2010fb}.  Unfortunately, not much can be inferred 
about dark matter (DM) beyond it's existence.  Some of the most popular 
candidates for dark matter are weakly interacting massive particles or WIMPs,
motivated by the WIMP(less) miracle \cite{Feng:2008ya}.
WIMPs naturally occur in models postulated to solve the gauge hierarchy problem.  Many theories have been found to 
address other issues in particle physics which also provide a viable dark matter
candidate.  Much effort has been put into understanding the predictions of these theoretically 
motivated models for experimental probes.  

We have many methods to search for dark matter.  These include astrophysical experiments which 
look for dark matter interacting with normal matter or dark matter annihilations, and collider searches
for new particles which may make up the missing dark matter.  However, understanding
implications one experiment has on another has largely been restricted to
placing limits, from one, on a favored theory and then deriving its predictions on the remaining 
parameter space to compare to other experiments \cite{Birkedal:2004xn,Beltran:2008xg,Cao:2009uw,Beltran:2010ww,Shepherd:2009sa}.  While this leaves us with different 
regions which are ``favored" by different models, it does not provide a good guide as to 
where we should be looking with future experiments in light of current results.  
Thus, it would be nice to establish a language for comparison between experiments.

Direct detection experiments search for dark matter particles recoiling off of nuclei as they 
travel through the Earth, while hadron colliders have the potential to produce dark matter from 
high energy hadronic material.  Both types of experiment probe the couplings of dark matter to hadronic
particles, and should complement each other, assuming the dark matter is light enough to be produced at colliders.  Many recent direct detection experiments have reported interesting, though,
conflicting results.  These include, CoGeNT \cite{Aalseth:2010vx,Aalseth:2011wp}, DAMA\cite{Bernabei:2010mq},
Xenon \cite{Aprile:2010um} and CDMS \cite{Ahmed:2010wy} and, in fact, models to reconcile these discrepancies  largely focus on the possibility of a light dark matter candidate \cite{Petriello:2008jj,Feng:2008dz,Schwetz:2011xm,Fox:2011px,Farina:2011pw,Feng:2011vu,Chang:2010yk,Kang:2011wb,DelNobile:2011je,Gao:2011ka,Frandsen:2011ts,Gao:2011bq,Chen:2011vd}.  

In previous work, \cite{Goodman:2010yf, Goodman:2010ku, 
Bai:2010hh},  the overlap between direct detection experiments and colliders was considered in terms of
effective four-point interactions between DM and hadronic particles, either quarks or
gluons. In our current work, we drop the assumption that the intermediate state is
heavy compared to the scales of interest and consider the simplest UV complete
models of DM interactions which could give rise to signals in direct detection
experiments and hadron colliders.  This was also considered for some special
cases in \cite{Bai:2010hh}.

This paper is organized as follows: In section\,\ref{sec:int} we discuss the
relevant interactions between dark matter and quarks at the renormalizable level,
in section\,\ref{sec:collider} we derive bounds on those interactions based on null
collider searches for monojets and missing energy, in section\,\ref{sec:direct} we
discuss the impact on direct detection parameters, and in section\,\ref{sec:conc} we conclude.

\section{Interactions of DM with quarks}
\label{sec:int}

In this article we work under the assumption that dark matter is composed of
one species of Standard Model singlet Dirac fermion, $\chi$ \cite{Burgess:2000yq}. 
Other candidates are possible, but the modification
of previous results due to the introduction of collider-accessible mediators
does not depend sensitively on the spin of the dark matter; it is dominated by
phase space and kinematical effects.

We consider all tree-level renormalizable interactions of DM particles, $\chi$,
with quarks which contribute to direct detection signals in the limit of zero
momentum transfer. This generically requires positing a new mediator particle, $\phi$,
whose quantum numbers depend on its particular interaction with quarks
and $\chi$.  We consider s-channel operators with a standard model singlet mediator
and t-channel operators with colored mediators. The channel of a process is defined by
the momentum combination which appears in the propagator of the mediating particle, either
the sum of initial state momenta for s-channel or the difference of one initial and one
final state momentum for t-channel. The mediators of t-channel models are color triplets with
electric charge of 2/3 or -1/3, such that they couple to up- and down-type quarks, respectively and 
we will assume, for simplicity, that they have identical masses. 
We do not assign specific $SU(2)_L$ charges to mediators since QCD dynamics or direct detection are dependent on parity, rather than chirality. As electroweak effects are subdominant to
QCD effects for these charged mediators, this is a good approximation for our study.

We assume that the dark matter
abundance is protected by a $Z_2$ or larger symmetry which implies that t-channel mediators are also
charged under this group.  Table I shows the labels
we assign to each model of dark matter interaction vertex; the first letter indicates the type 
of dark matter, in this case Dirac WIMPs, and the second letter indicates the
DM annihilation channel, either s or t.  These interaction vertices also contribute to indirect detection \cite{Cheung:2010ua}.
Similar interactions with leptons can also be considered \cite{Mambrini:2011pw}.  While they have
interesting motivations in astrophysical observations, their couplings
do not significantly contribute to either hadron collider signals or direct
detection experiments and so we will not consider them here.  It should be noted that one can also consider couplings
to other Standard Model particle, e.g. \cite{Kanemura:2010sh}, as well.

The s-channel models are assumed to be flavor diagonal in couplings and
therefore are not significantly constrained by precision measurements. The
t-channel models can, in principle, contribute to FCNC decays of hadrons at one
loop order, but, we will assume that these decays are suppressed by some
mechanism and not concern ourselves
further with this. The couplings of DS1 are scaled by the quark Yukawa coupling
appropriate to ensure Minimal Flavor Violation \cite{Buras:2000dm}.\footnote[1]{The flavor phenomenology of dark
matter models has been more thoroughly considered in \cite{Kamenik:2011nb,Cheung:2010zf}.}  We thus have, in all posited
models of DM interactions, three parameters, the DM mass, $M_\chi$, the mediator
mass, $M_\phi$, and the coupling, $g$. In principle the couplings in s-channel
models to DM and to quarks can be unrelated, but one can only bound
the product of the two couplings at colliders and direct detection experiments,
so we take them to be identical.

Integrating out the mediator $\phi$, leads to contact interactions which are suppressed by a scale $M_*\equiv M_\phi/g$, which relates to models considered in \cite{Goodman:2010ku,Goodman:2010yf,Goodman:2010qn,Rajaraman:2011wf}.  In the case of s-channel mediators these operators are identical to those discussed previously which contribute to direct detection.  One can use Fierz transformations to write the t-channel models as linear combinations of models previously considered.  

\begin{table}[h]
 \hspace{0.033\textwidth}
\label{tab:caption0}
 \begin{minipage}{0.4\textwidth}
  \centering
 \begin{tabular}{|c|c|c|}
\hline
Name  & $\chi$ Operator & $q$ Operator \\
\hline
DS1 & $g\bar{\chi}\chi\phi$ & $g\lambda_q\bar{q}q\phi$ \\
DS2 & $g\bar{\chi}\gamma^\mu\chi\phi_\mu$ & $g\bar{q}\gamma^\mu q\phi_\mu$ \\
DS3 & $g\bar{\chi}\gamma^\mu\gamma^5\chi\phi_\mu$ & $g\bar{q}\gamma^\mu\gamma^5 q\phi_\mu$ \\
\hline
\end{tabular}
  \label{tab:caption1}
 \end{minipage}
\hspace{0.033\textwidth}
\begin{minipage}{0.4\textwidth}
\centering
\begin{tabular}{|c|c|}
\hline
Name & Operator \\
\hline
DT1 & $g\bar{q}\chi\phi$ \\
DT2 & $g\bar{q}\gamma^5\chi\phi$ \\
\hline
\end{tabular}
\label{tab:caption2}
\end{minipage}
\caption{\textnormal{Models of DM coupling to quarks. The models are named such that the first letter describes the DM particle as being Dirac fermions and the second letter gives what channel DM annihilations proceed through.}}
\end{table}

\section{Collider limits and reaches}
\label{sec:collider}

We derive limits and future reaches for collider searches looking for missing
energy signatures.
For all the bounds which we generate from colliders we only consider theories which
may possibly be perturbative, that is, models which have $\alpha\equiv g^2/4\pi<4\pi$. We consider
any model which has a greater coupling to not be the correct picture for
understanding the dynamics of dark matter interactions. Note that, while some interactions of the
model DS1 remain perturbative for larger values of $g$ due to the presence of the quark Yukawa
coupling, the coupling to top quarks, which is required by our hypothesis of flavor structure, becomes
non-perturbative according to this criterion at approximately the same point. We thus assume perturbativity,
as defined above, to be a theoretical bound on all models considered and we report only
experimental bounds which are more stringent than this.

The simplest signature of an event which contains dark matter visible in direct
detection is one of a jet and missing energy. While additional jets are possible,
they generally require a more involved analysis,
so we focus only on the monojet case here.\footnote[2]{Other final states are possible and this has been explored in
\cite{Wang:2011sx}.}  In the context of the t-channel mediators
this amounts to searching for associated production of a mediator and a dark matter
candidate, rather than pair production of the mediators. In the special case of
Supersymmetry, this process has been considered at NLO in\,\cite{Binoth:2011xi}.  A treatment similar to ours,
using simplified models to interpret multijet and missing energy searches, has recently been published by
the ATLAS collaboration\,\cite{AtlasConf2011}.

Our signal events were simulated using MadGraph and MadEvent version 4.4.56 \cite{Alwall:2007st},
using CTEQ6l PDFs with the renormalization scale chosen to be $\mu^2=M_\chi^2+P_{T,j}^2$.
The events were hadronized using Pythia $6.2$  \cite{Sjostrand:2006za}, and detector
simulation was done through PGS$4$ \cite{PGS}.

We utilize the recent search for monojets and missing energy, by the ATLAS collaboration \cite{AtlasConf2011}, to derive bounds on the coupling strength for each choice of dark matter mass and the mediator mass. The search includes three different sets of selection criteria which they call LowPt, HighPt, and VeryHighPt. The cuts required of each are shown in table II, along with the 95\% CL limit which was generated on the effective cross section from new physics for each selection. 
\begin{table}[h]
\begin{tabular}{|c|c|c|c|c|c|}
\hline
Name&$\met$&monojet $P_t$&2nd jet $P_t$&3rd jet $P_t$&95\% limit\\
\hline
LowPt&$>120$ GeV&$>120$ GeV&$<30$ GeV&$<30$ GeV&$1.7$ pb\\
HighPt&$>220$ GeV&$>250$ GeV&$<60$ GeV&$<30$ GeV&$0.11$ pb\\
VeryHighPt&$>350$ GeV&$>300$ GeV&$<60$ GeV&$<30$ GeV&$0.035$ pb\\
\hline
\end{tabular} 
\caption{Cuts used in ATLAS monojet plus missing energy search \cite{AtlasConf2011}.}
\end{table}

\begin{figure}[ht]
  \begin{center}    
      \includegraphics[width=.47\textwidth]{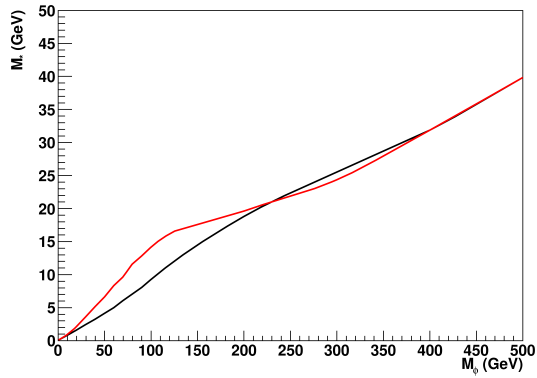}    
  \end{center}
  \caption{\label{fig:DS1} Bounds on effective interaction strength $M_*=M_\phi/g$ for the operator DS1. Bounds are presented only for massless $\chi$, as all other bounds are weaker than our adopted perturbativity limit of $g<4\pi$. The red curve shows bounds resulting from the VeryHighPt analysis, and the black curve shows those resulting from the LowPt analysis.  }
  \end{figure}

\begin{figure}[ht]
  \begin{center}
    \subfigure{
      \includegraphics[width=.47\textwidth]{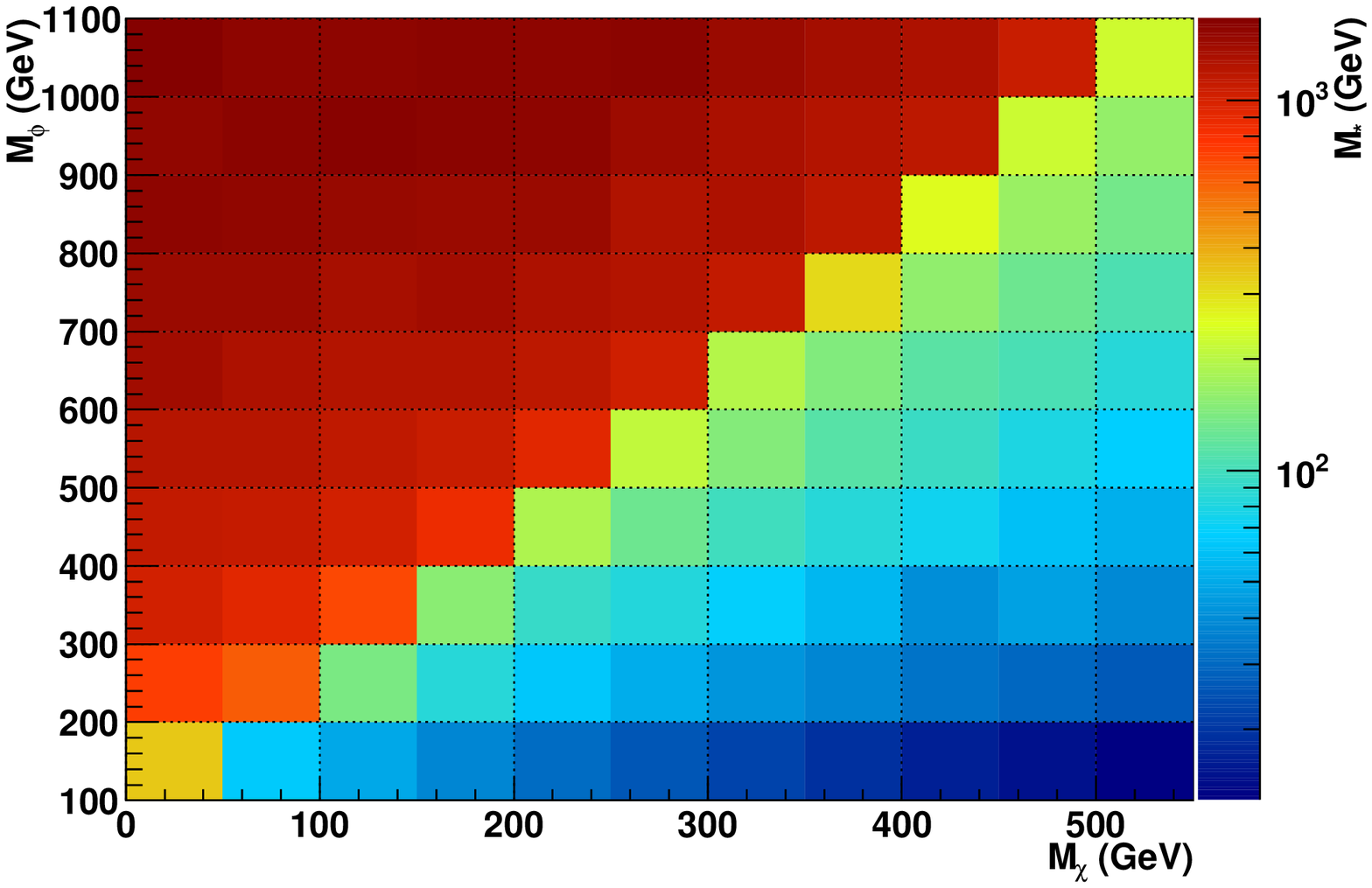}
    }
    \subfigure{
      \includegraphics[width=.47\textwidth]{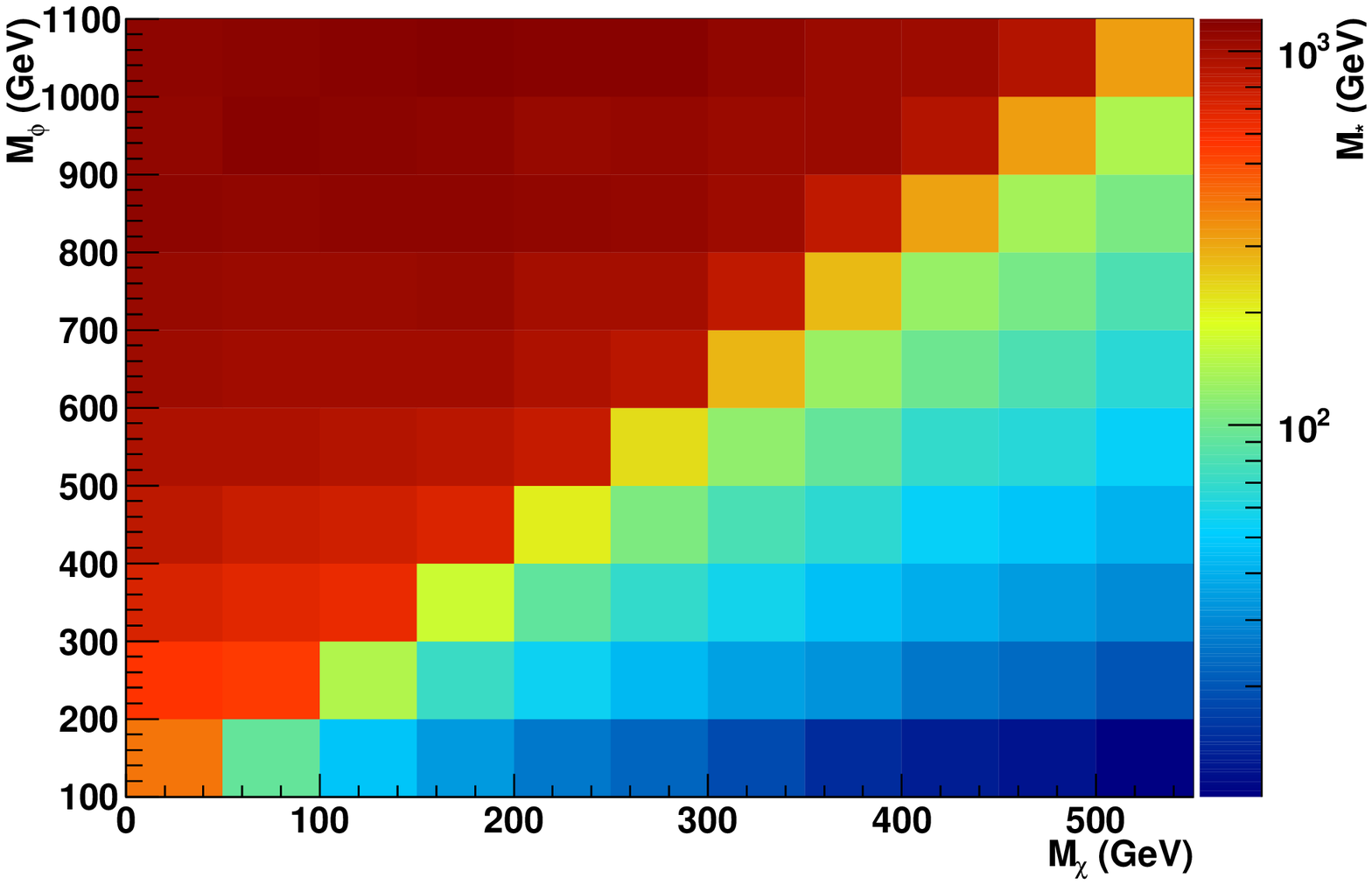}
    }
    \subfigure{
      \includegraphics[width=.47\textwidth]{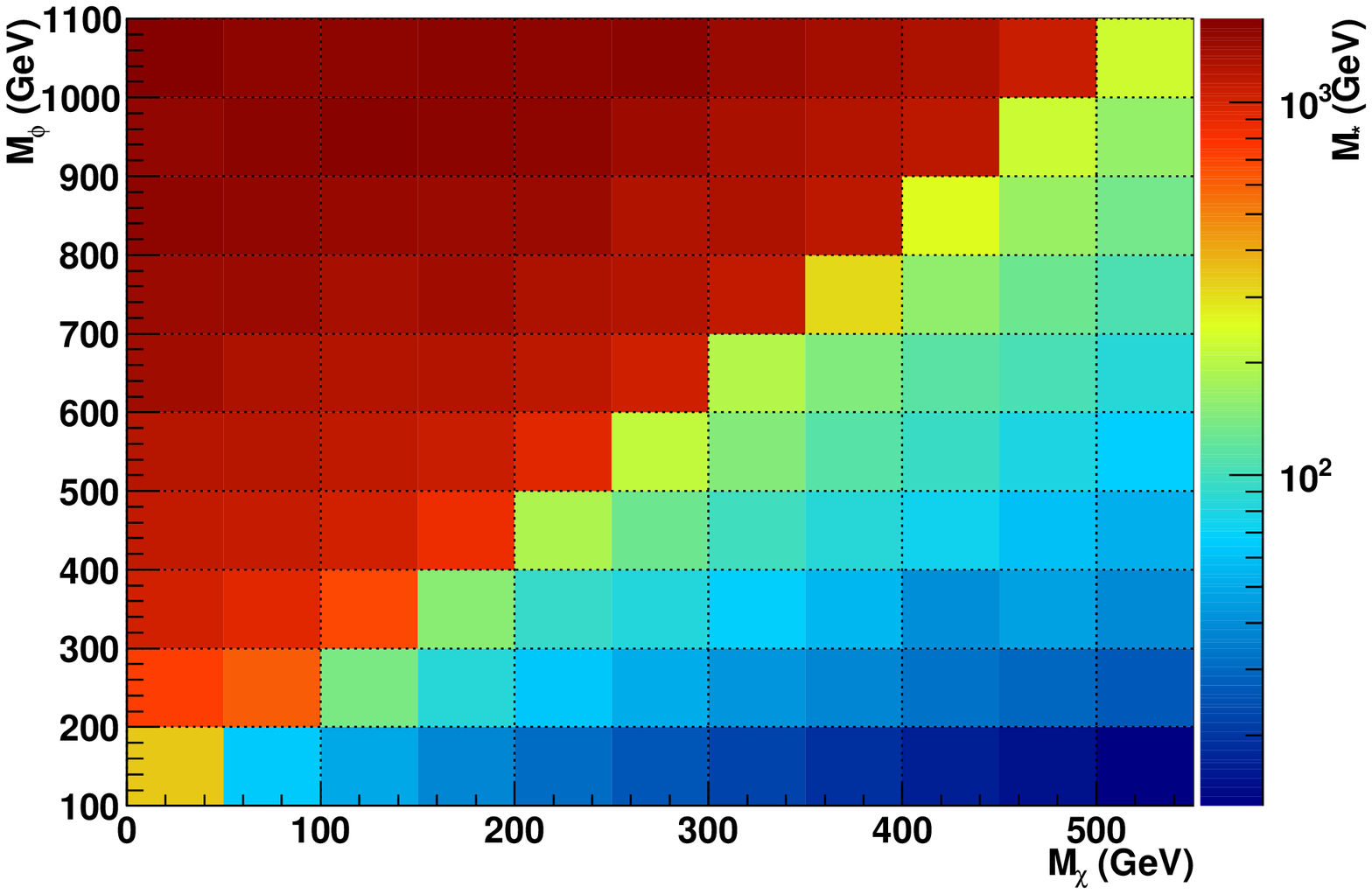}
    }
    \subfigure{
      \includegraphics[width=.47\textwidth]{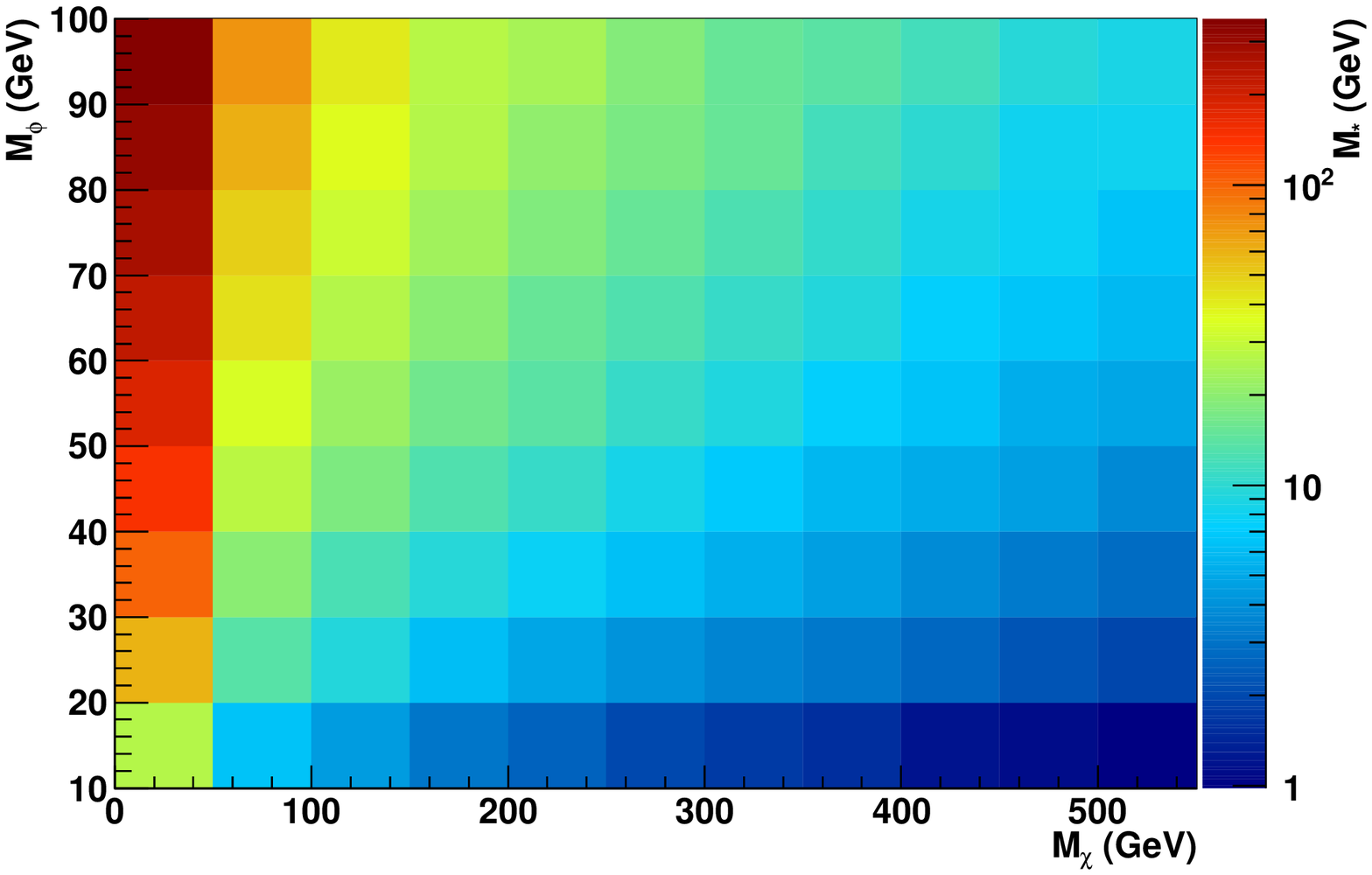}
    }
  \end{center}
  \caption{\label{fig:DS5} Bounds on effective interaction strength $M_*=M_\phi/g$ for the operator DS2. Note that the perturbativity constraint of $g<4\pi$ replaces bounds weaker than that constraint. The left figures show bounds resulting from the VeryHighPt analysis, and the right figures show those resulting from the LowPt analysis.  }
  \end{figure}

\begin{figure}[ht]
  \begin{center}
    \subfigure{
      \includegraphics[width=.47\textwidth]{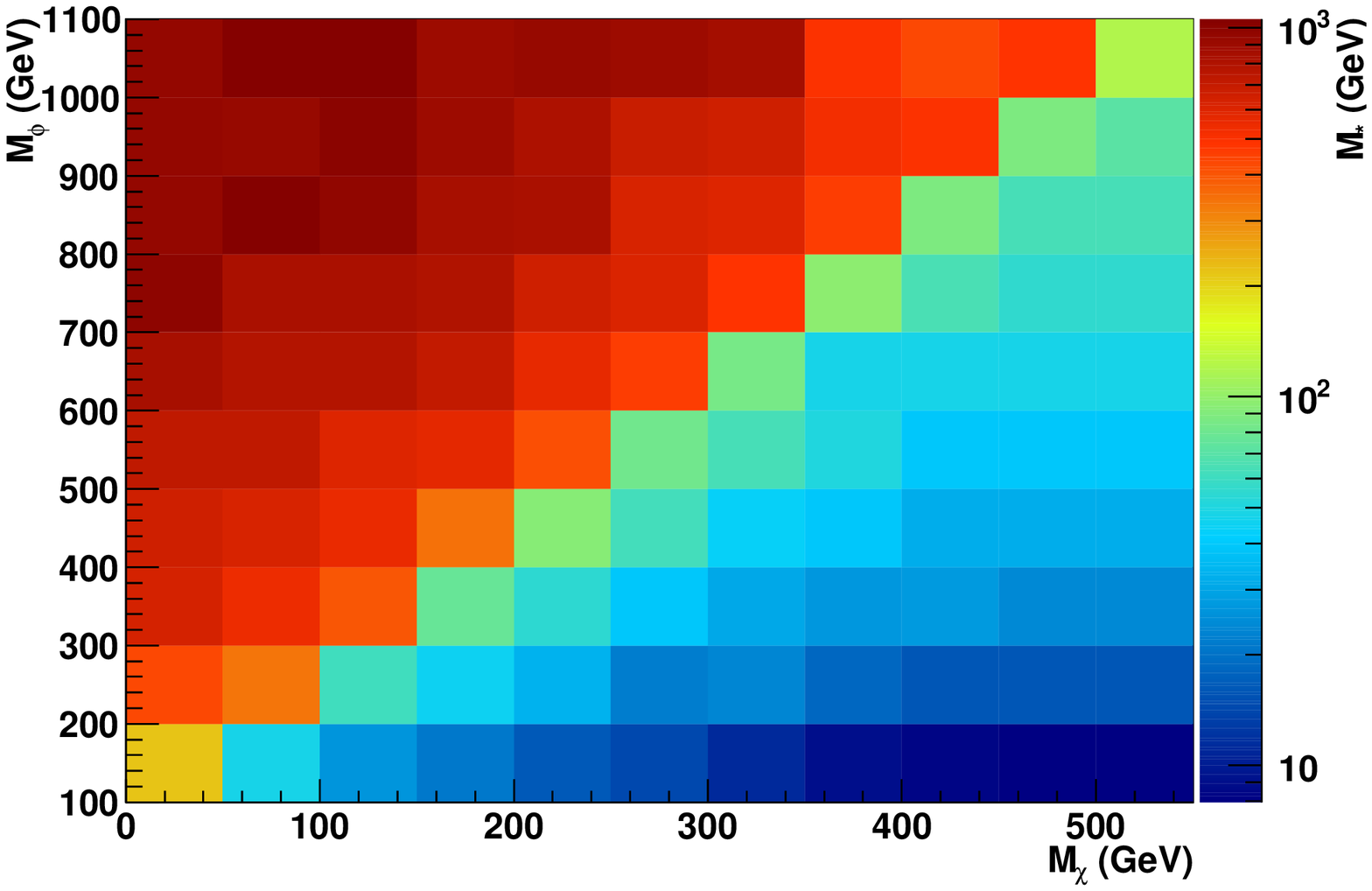}
    }
    \subfigure{
      \includegraphics[width=.47\textwidth]{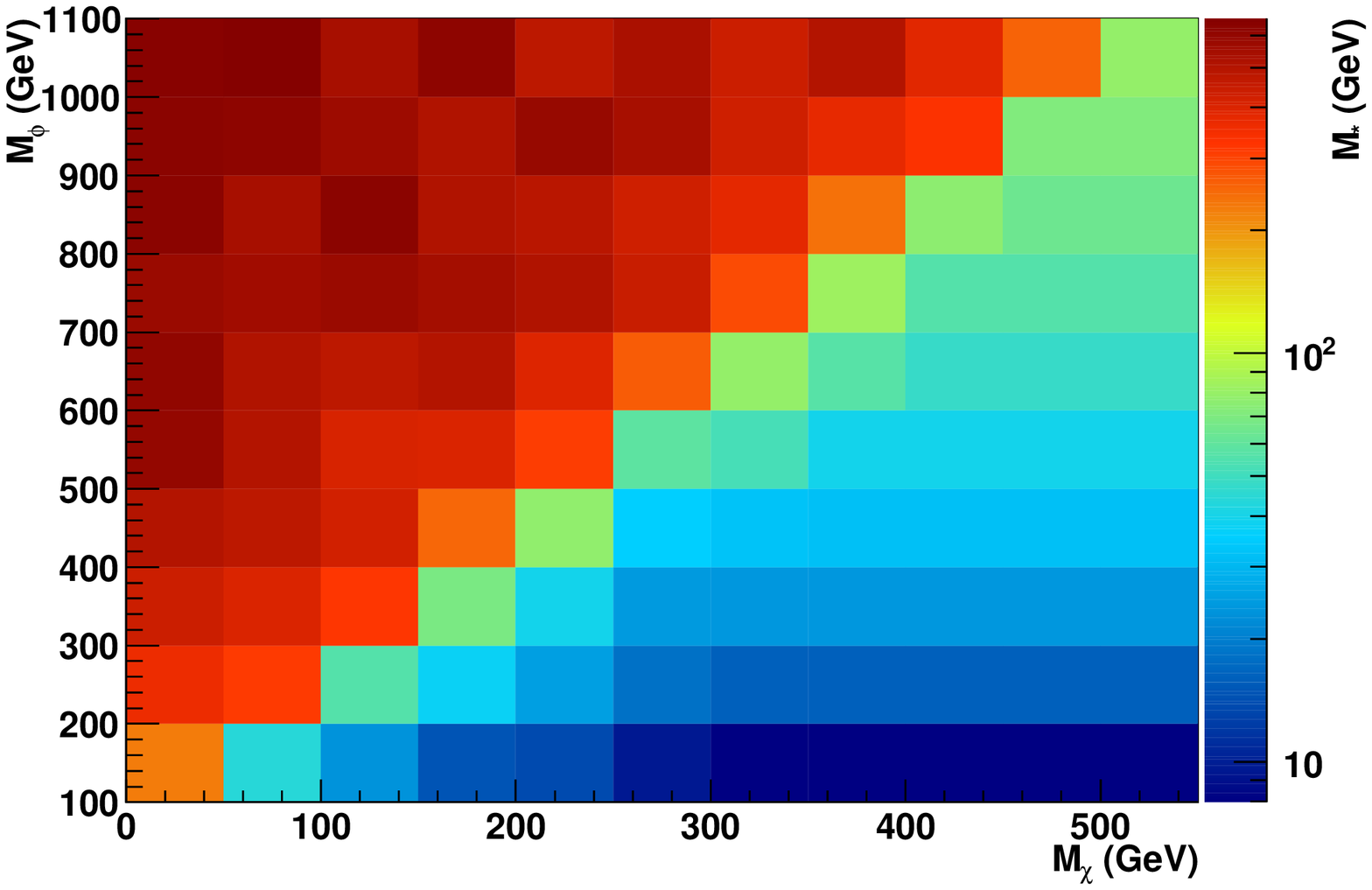}
    }
    \subfigure{
      \includegraphics[width=.47\textwidth]{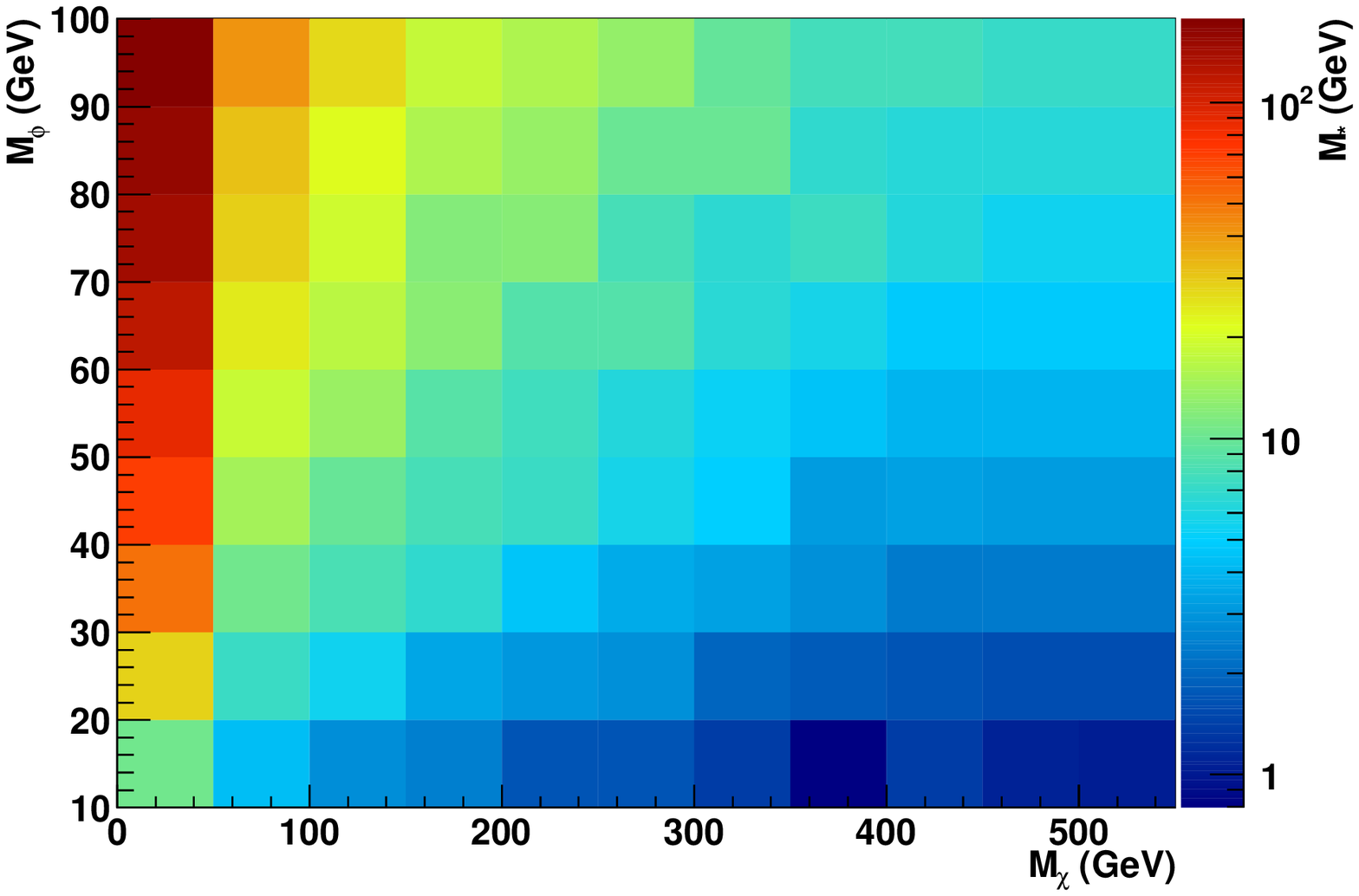}
    }
    \subfigure{
      \includegraphics[width=.47\textwidth]{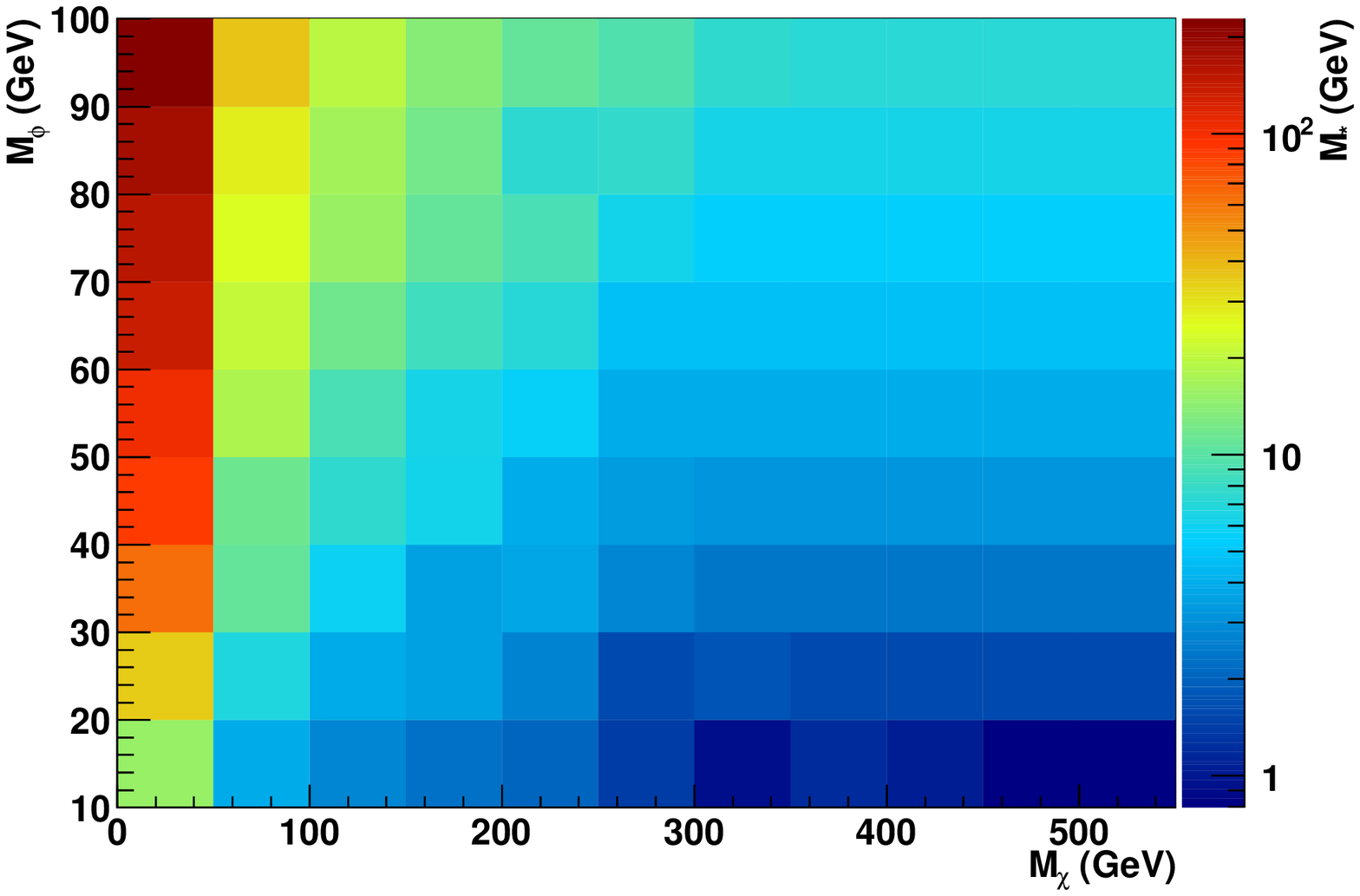}
    }
  \end{center}
  \caption{\label{fig:DS8} Bounds on effective interaction strength $M_*=M_\phi/g$ for the operator DS3. Note that the perturbativity constraint of $g<4\pi$ replaces bounds weaker than that constraint. The left figures show bounds resulting from the VeryHighPt analysis, and the right figures show those resulting from the LowPt analysis.  }
  \end{figure}

\begin{figure}[ht]
  \begin{center}
    \subfigure{
      \includegraphics[width=.47\textwidth]{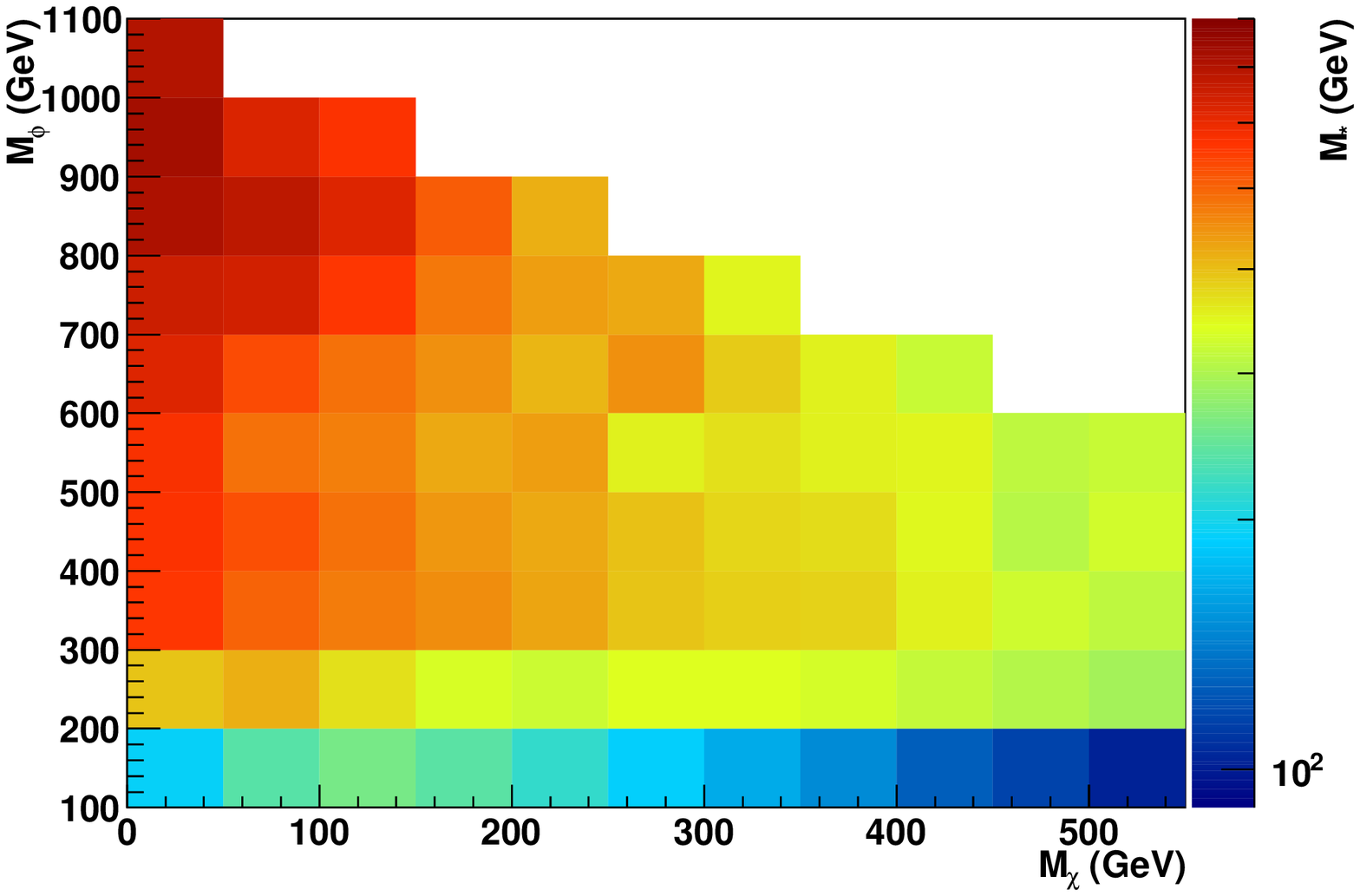}
    }
    \subfigure{
      \includegraphics[width=.47\textwidth]{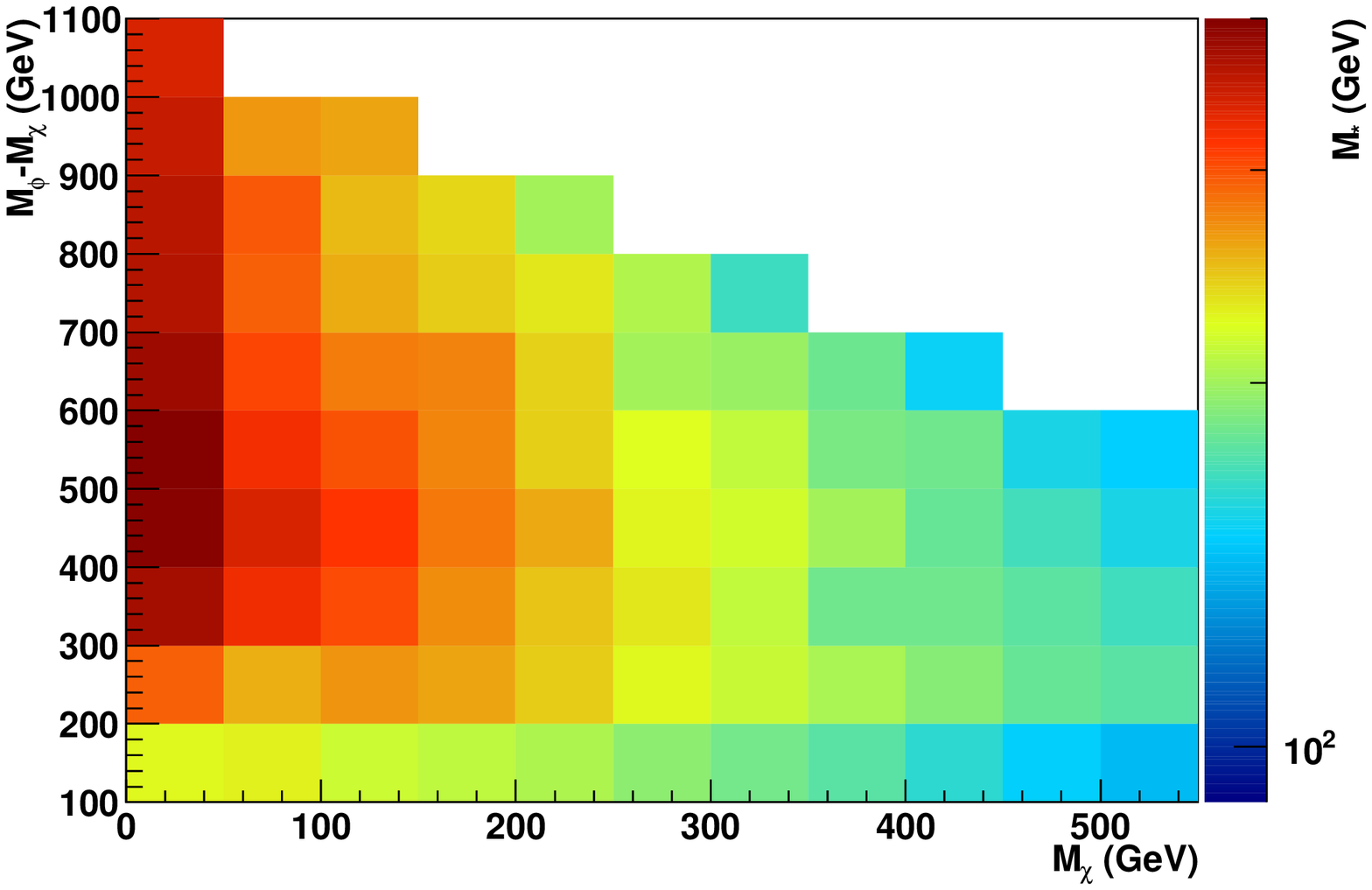}
    }
    \subfigure{
      \includegraphics[width=.47\textwidth]{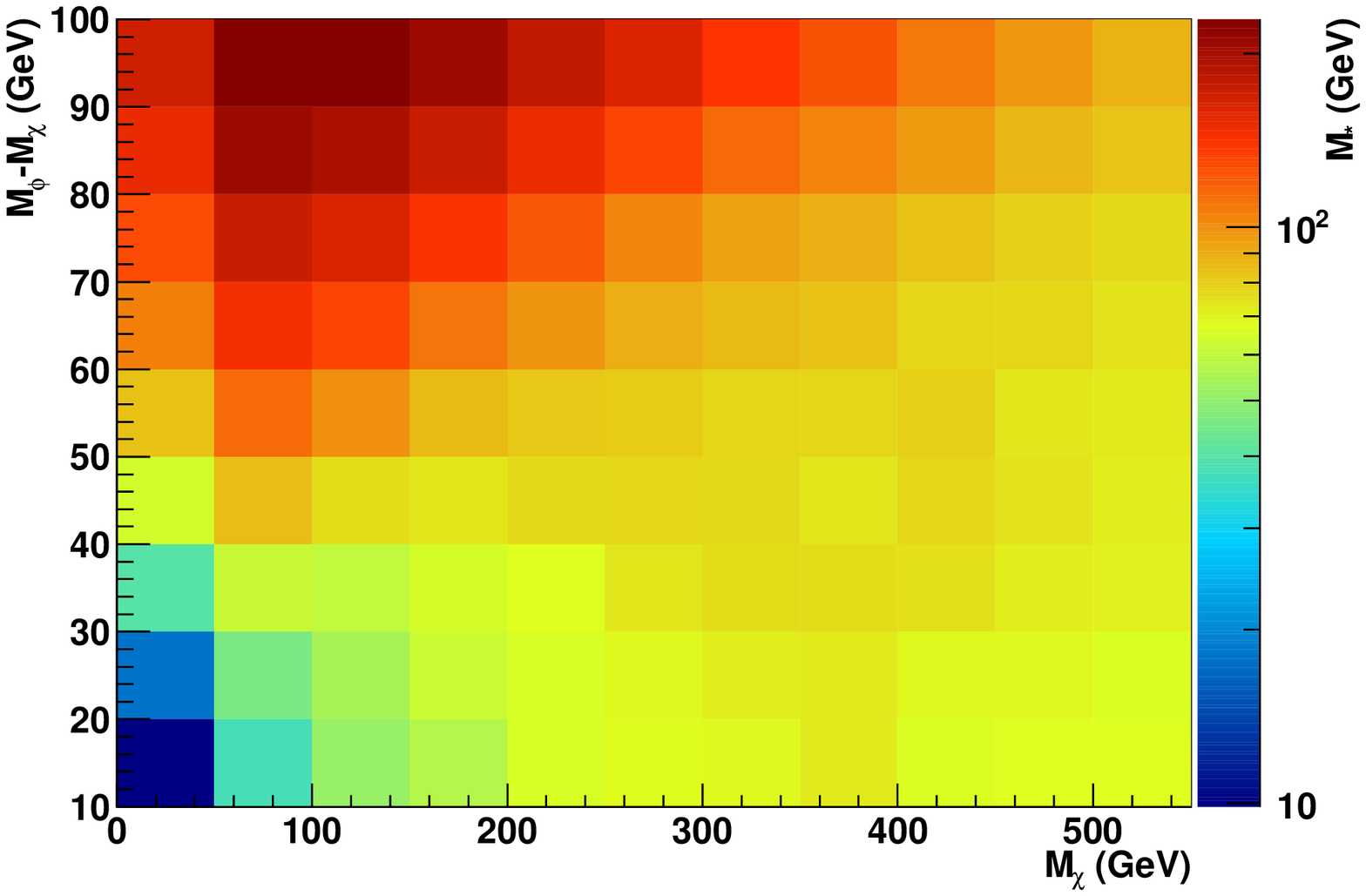}
    }
    \subfigure{
      \includegraphics[width=.47\textwidth]{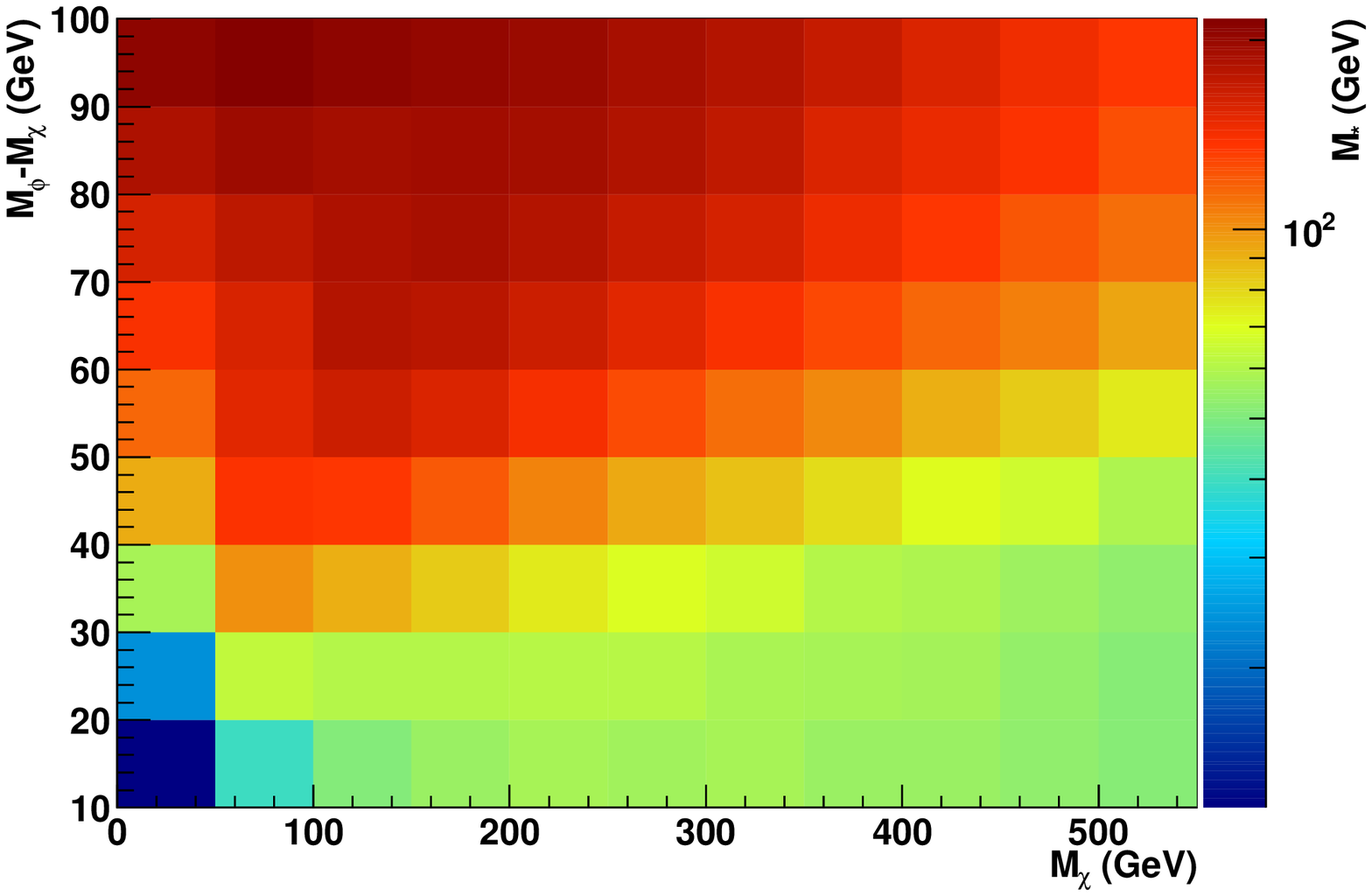}
    }
  \end{center}
  \caption{\label{fig:DT1} Bounds on effective interaction strength $M_*=M_\phi/g$ for the operator DT1. Note that the perturbativity constraint of $g<4\pi$ replaces bounds weaker than that constraint. The left figures show bounds resulting from the VeryHighPt analysis, and the right figures show those resulting from the LowPt analysis.  }
  \end{figure}

\begin{figure}[ht]
  \begin{center}
    \subfigure{
      \includegraphics[width=.47\textwidth]{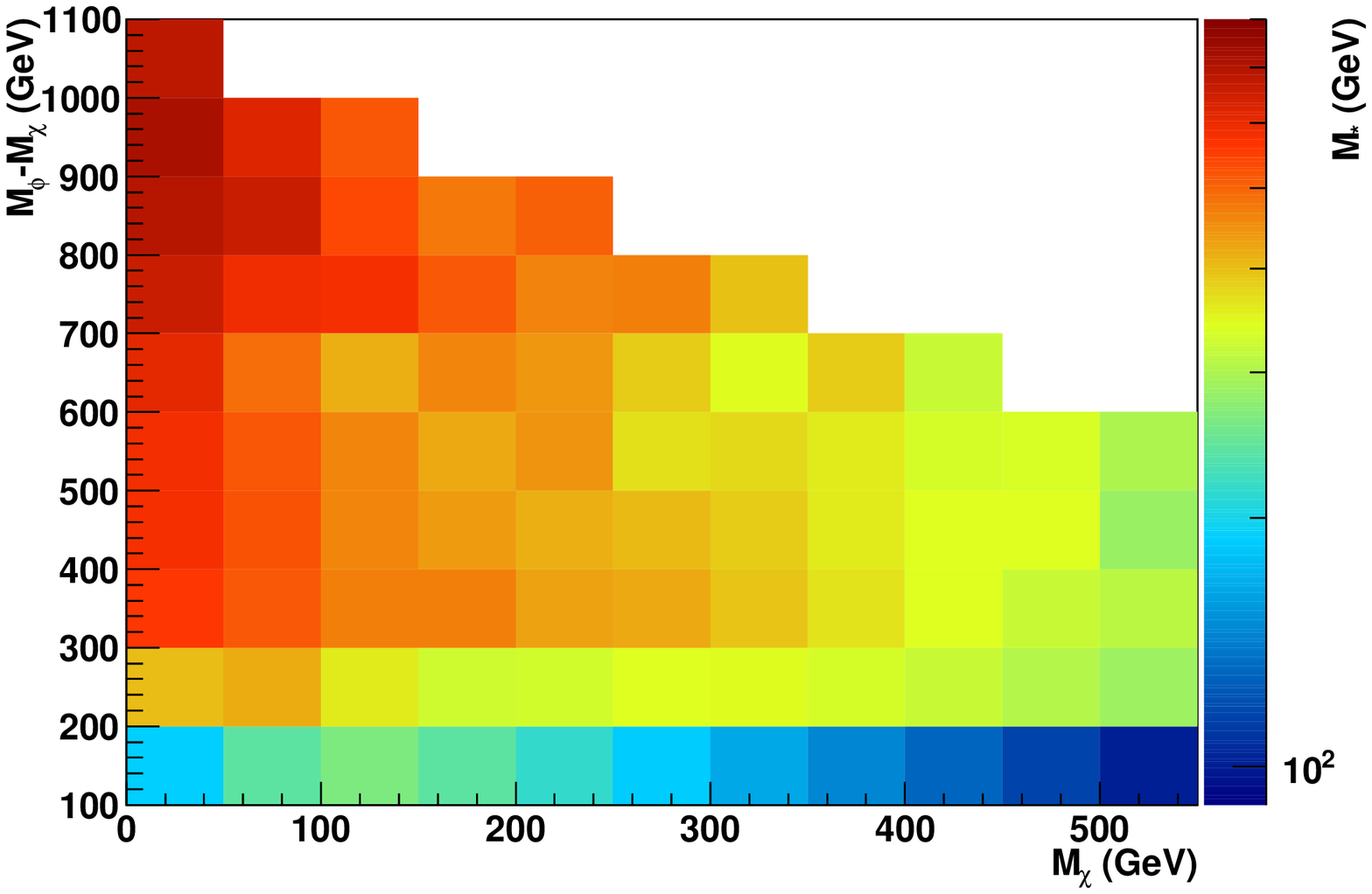}
    }
    \subfigure{
      \includegraphics[width=.47\textwidth]{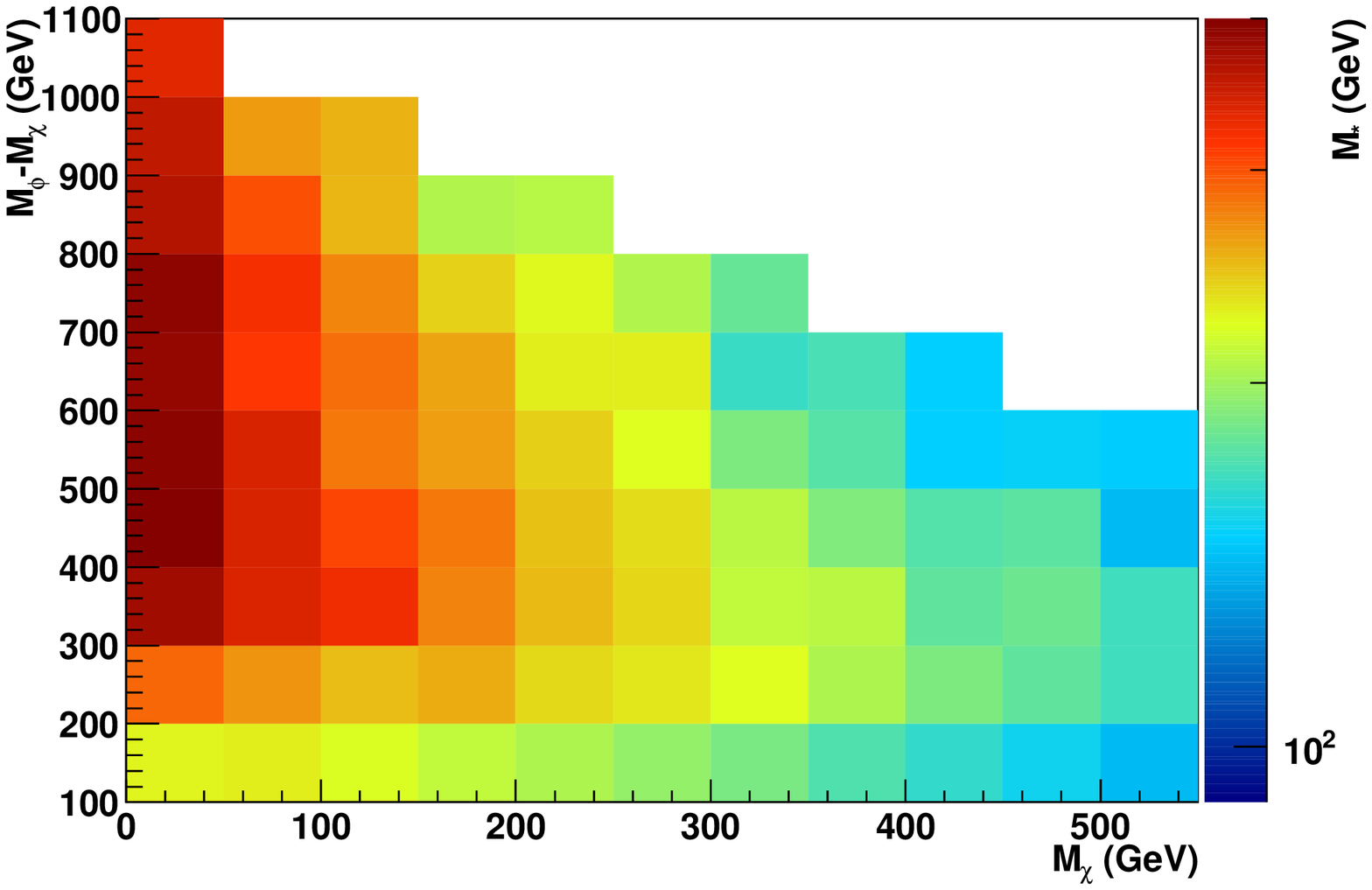}
    }
    \subfigure{
      \includegraphics[width=.47\textwidth]{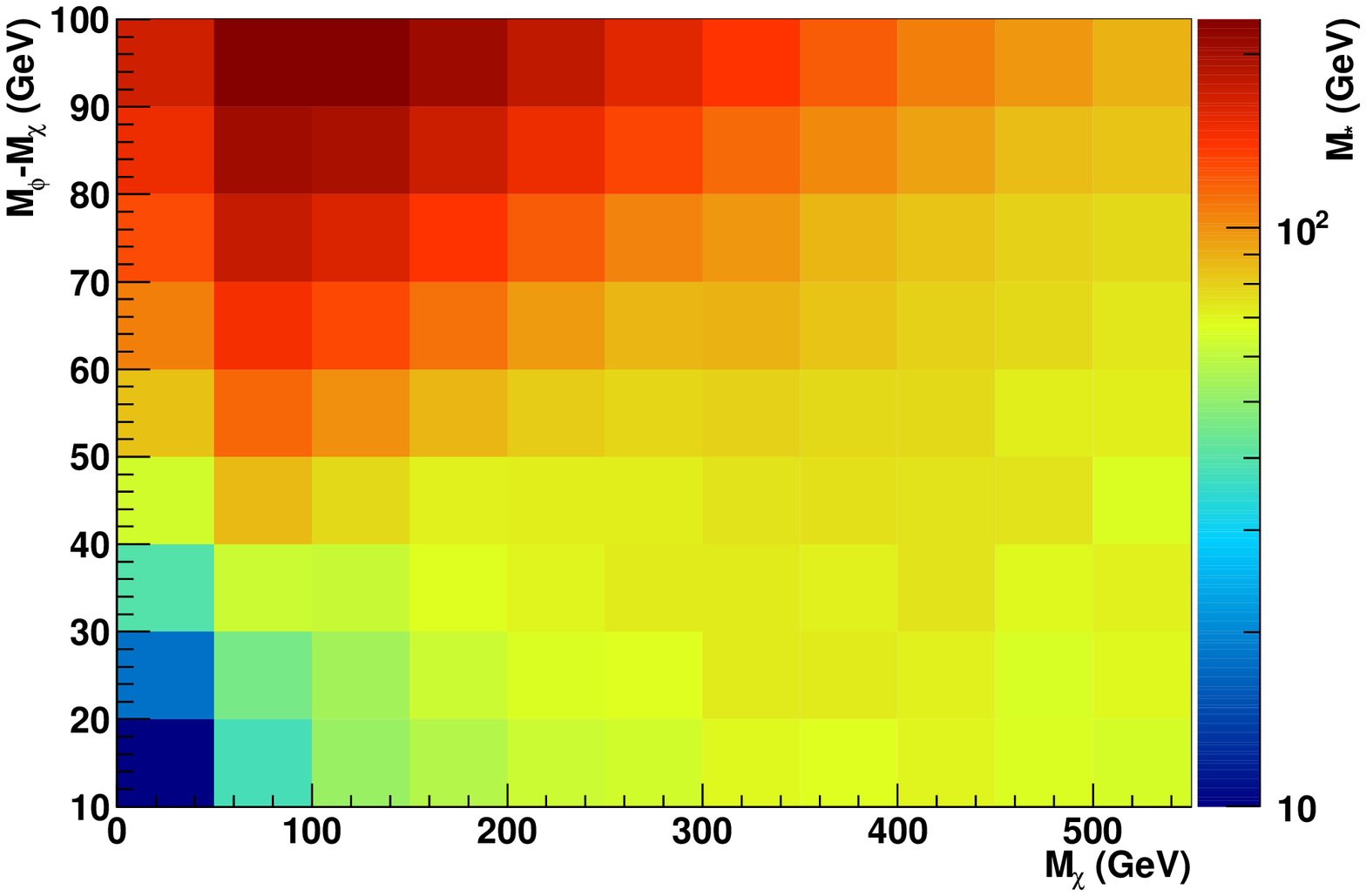}
    }
    \subfigure{
      \includegraphics[width=.47\textwidth]{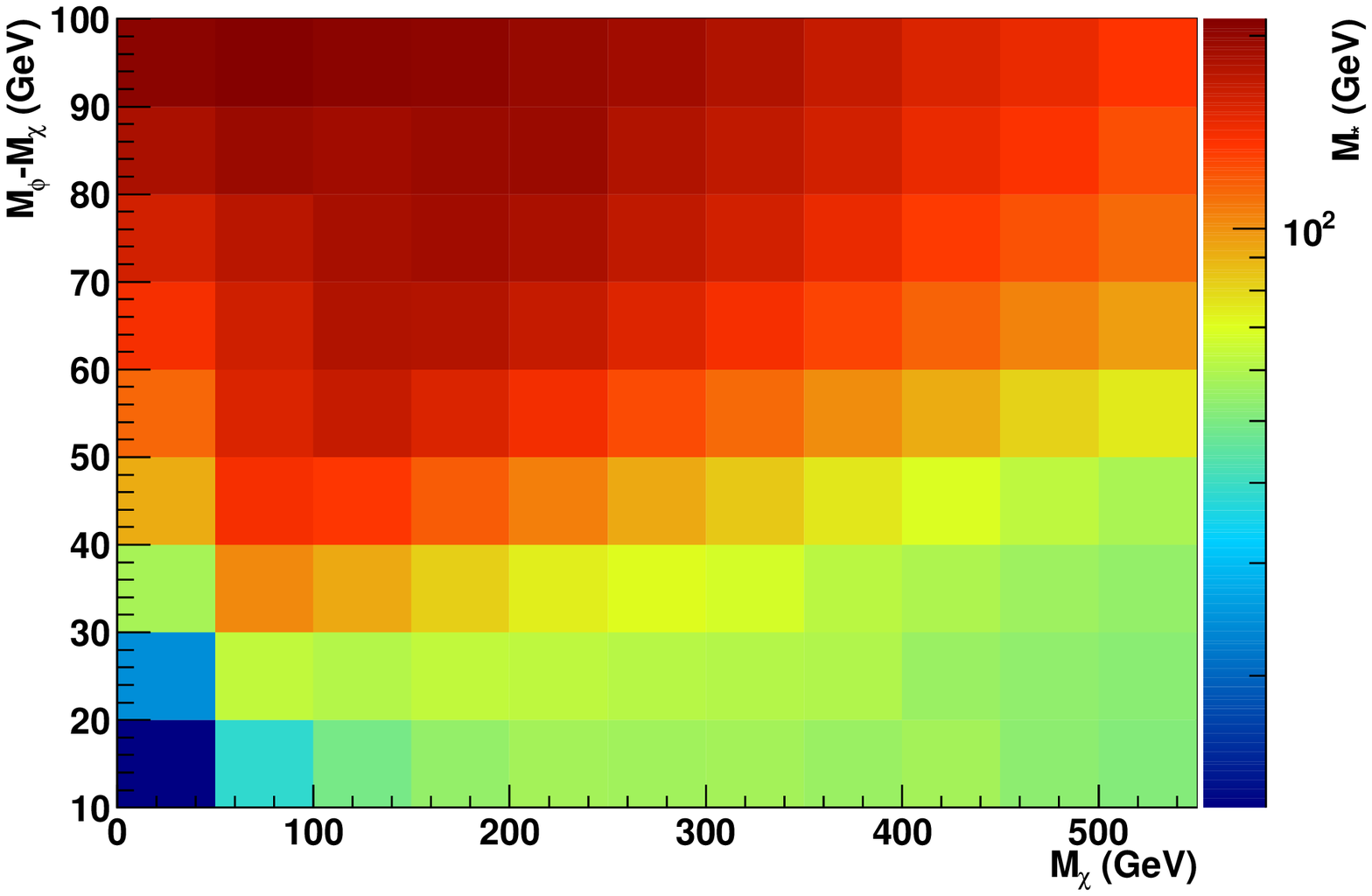}
    }
  \end{center}
  \caption{\label{fig:DT2} Bounds on effective interaction strength $M_*=M_\phi/g$ for the operator DT2. Note that the perturbativity constraint of $g<4\pi$ replaces bounds weaker than that constraint. The left figures show bounds resulting from the VeryHighPt analysis, and the right figures show those resulting from the LowPt analysis.  }
  \end{figure}

We considered specifically the LowPt and VeryHighPt criteria, applying those cuts to our generated
signal events and then finding the necessary coupling strength to saturate the bounds quoted above.
We present bounds on $M_{*}\equiv M_\phi/g$ from the effective theory previously considered in\,\cite{Goodman:2010ku,Goodman:2010yf,Goodman:2010qn,Rajaraman:2011wf}, as we are interested in the change of bounds from previous results due to the introduction of collider-accessible mediators.
The resulting 95\% CL limits are presented in figures\,\ref{fig:DS1}-\ref{fig:DT2}. We note that, while bounds are generically weakened when mediators are lighter than the characteristic scale of the process which they are involved in ($M_\phi\lsim2M_\chi$ for s-channel, $M_\phi\lsim M_\chi+\met_{\rm cut}$ for t-channel), above those thresholds the bounds are stronger or comparable to those previously derived within the assumption of a collider-inaccessible mediator.


\section{Direct Detection of Candidate Models}
\label{sec:direct}

The direct detection calculations for s-channel annihilation models are rather
straightforward, as the mediator, even if very light by other standards, is
still very massive compared to the scales involved in direct detection, which
are typically of order $\sim$ keV-MeV. We then can integrate out the mediator
and return to the calculations of \cite{Goodman:2010ku,Goodman:2010yf}, using nuclear
matrix elements from \cite{Giedt:2009mr,Belanger:2008}. 

For t-channel models, we again must assume we can
integrate out the mediator
particle, which now requires it to be heavier than the DM candidate. This
was already necessary to allow our assumed symmetry to protect the DM
abundance. Once we have integrated out the mediator we can apply the appropriate
Fierz transformations to find the contribution to direct detection scatterings
in the language of contact interactions. The direct detection
cross sections predicted by the various contributing models are

\bea
\sigma^{DS1}_{0,SI}&=&1.60\times10^{-37} {\rm cm}^2\times \left(\frac{\mu_\chi}{1 {\rm GeV}}\right)^2
\left(\frac{20 {\rm GeV}}{M_*}\right)^4\left(\frac{20 {\rm GeV}}{v}\right)^2,\\
\sigma_{0,SI}^{DS2}&=&1.38\times10^{-37} {\rm cm}^2\times \left(\frac{\mu_\chi}{1 {\rm GeV}}\right)^2
\left(\frac{300 {\rm GeV}}{M_*}\right)^4,\\
\sigma^{DS3}_{0,SD}&=&9.18\times10^{-40} {\rm cm}^2\times \left(\frac{\mu_\chi}{1 {\rm GeV}}\right)^2
\left(\frac{300 {\rm GeV}}{M_*}\right)^4,\\
\sigma^{DT1}_{0,SI}&=&2.21\times10^{-37} {\rm cm}^2\times \left(\frac{\mu_\chi}{1 {\rm GeV}}\right)^2
\left(\frac{300 {\rm GeV}}{M_*}\right)^4,\\
\sigma^{DT1}_{0,SD}&=&2.30\times10^{-40} {\rm cm}^2\times \left(\frac{\mu_\chi}{1 {\rm GeV}}\right)^2
\left(\frac{300 {\rm GeV}}{M_*}\right)^4,\\
\sigma^{DT2}_{0,SI}&=&8.08\times10^{-38} {\rm cm}^2\times \left(\frac{\mu_\chi}{1 {\rm GeV}}\right)^2
\left(\frac{300 {\rm GeV}}{M_*}\right)^4.
\eea

Of particular interest is the interference behavior giving rise to the cross sections due to t-channel models. Model DT1 has constructive interference between the Fierz products which contribute to spin-dependent scattering, while in model DT2 they cancel, giving no spin-dependent signal. The same is true for spin-independent scattering, though the cancellation is less effective so the cross section is merely reduced, not eliminated.

\begin{figure}[ht]
  \begin{center}
    \includegraphics[width=.9\textwidth]{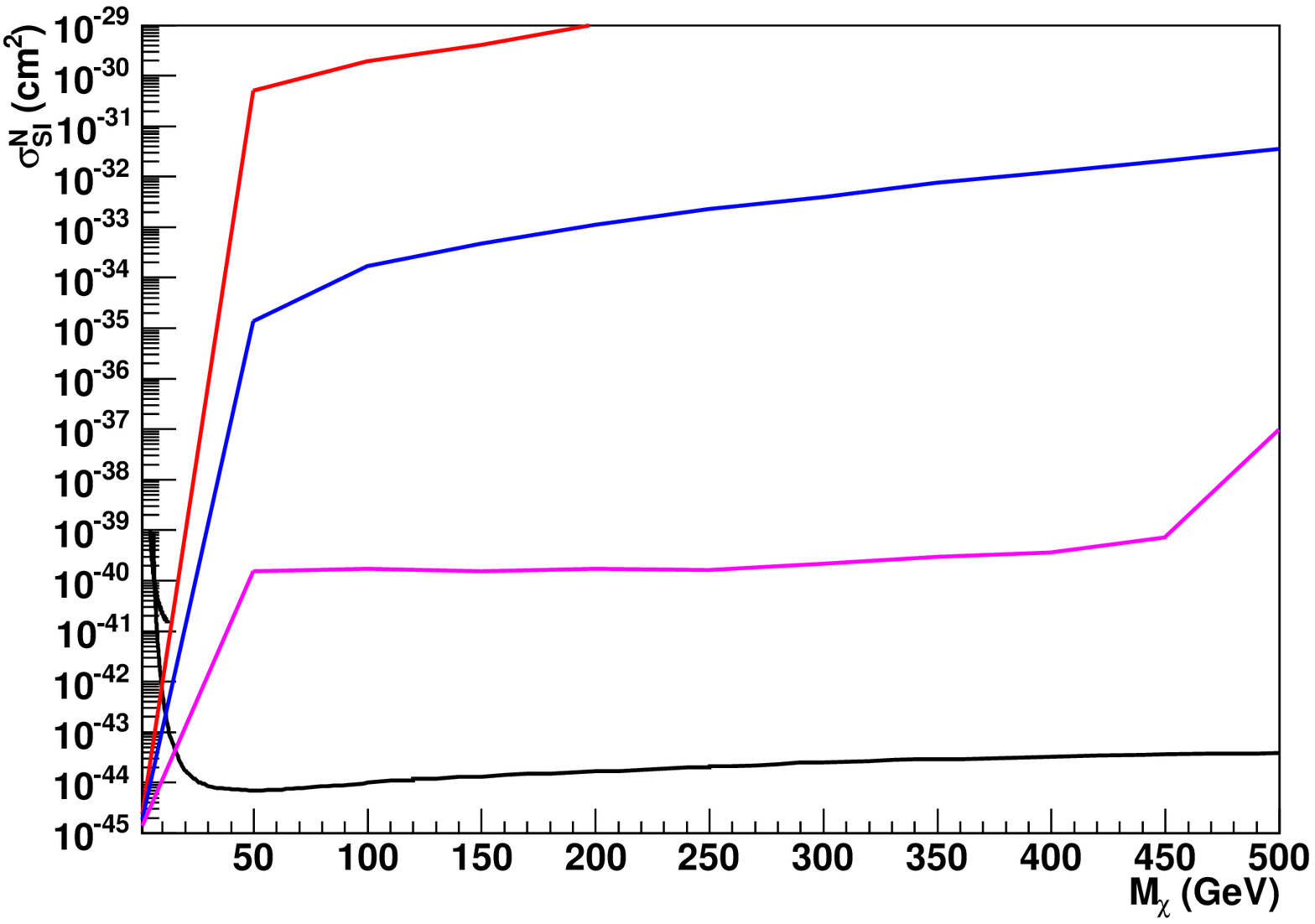}
  \end{center}
  \caption{\label{fig:sigDS5} Constraints on direct detection spin-independent cross section as a function of dark matter mass. The black curves correspond to the experimental limits from Xenon 100 and the CDMSII low-threshold analysis, and the red, blue, and magenta curves correspond to the bounds for operator DS2 with mediators of mass $M_\phi$ of 10, 100, and 1000 GeV, respectively.}
\end{figure}

\begin{figure}[ht]
  \begin{center}
    \includegraphics[width=.9\textwidth]{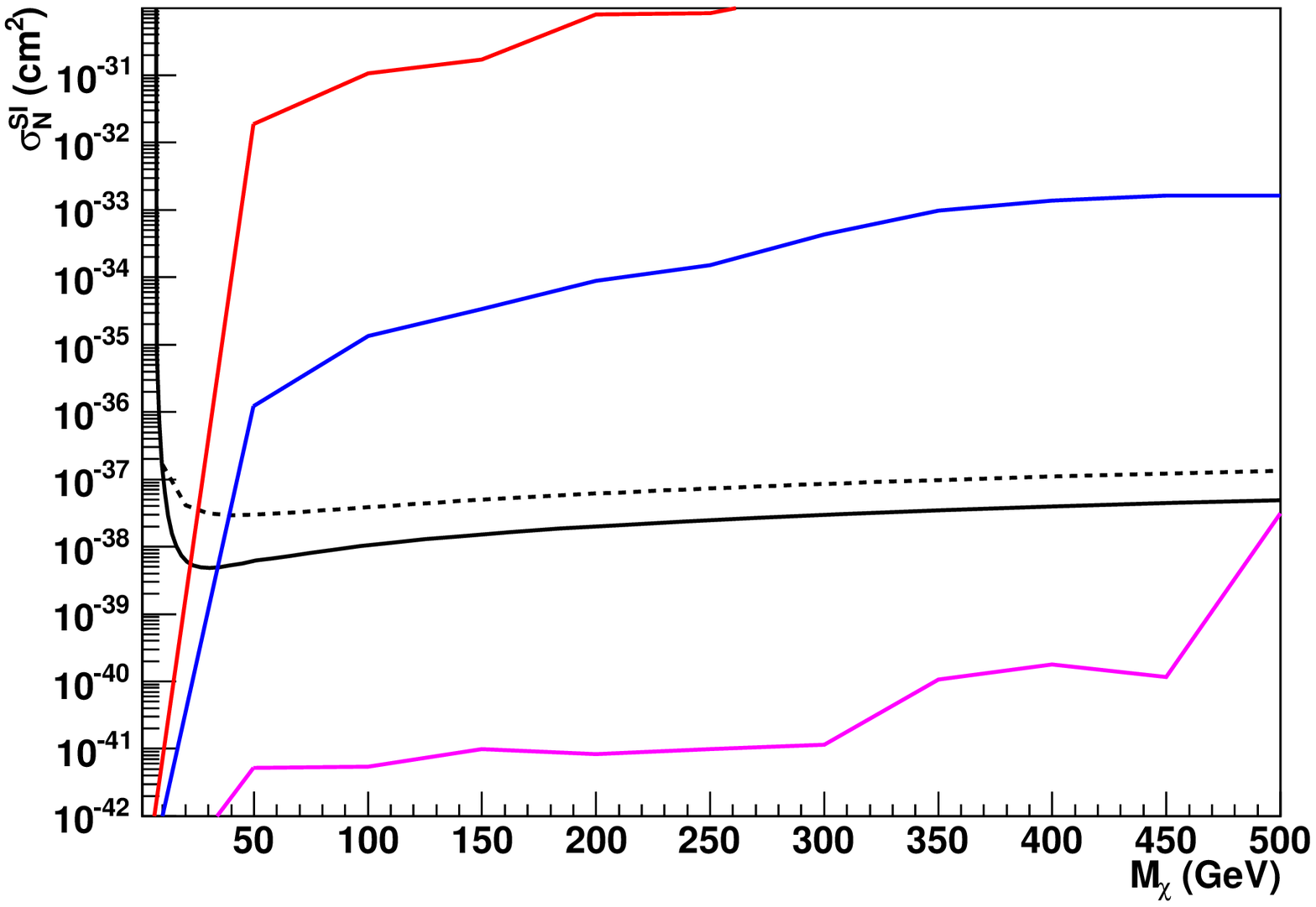}
  \end{center}
  \caption{\label{fig:sigDS8} Constraints on direct detection spin-dependent cross section as a function of dark matter mass. The black curves correspond to the experimental limits from Xenon 10 on neutron scatterings (solid curve) and PICASSO on proton scattering (dashed curve), and the red, blue, and magenta curves correspond to the bounds for operator DS3 with mediators of mass $M_\phi$ of 10, 100, and 1000 GeV, respectively.}
\end{figure}

\begin{figure}[ht]
  \begin{center}
    \includegraphics[width=.9\textwidth]{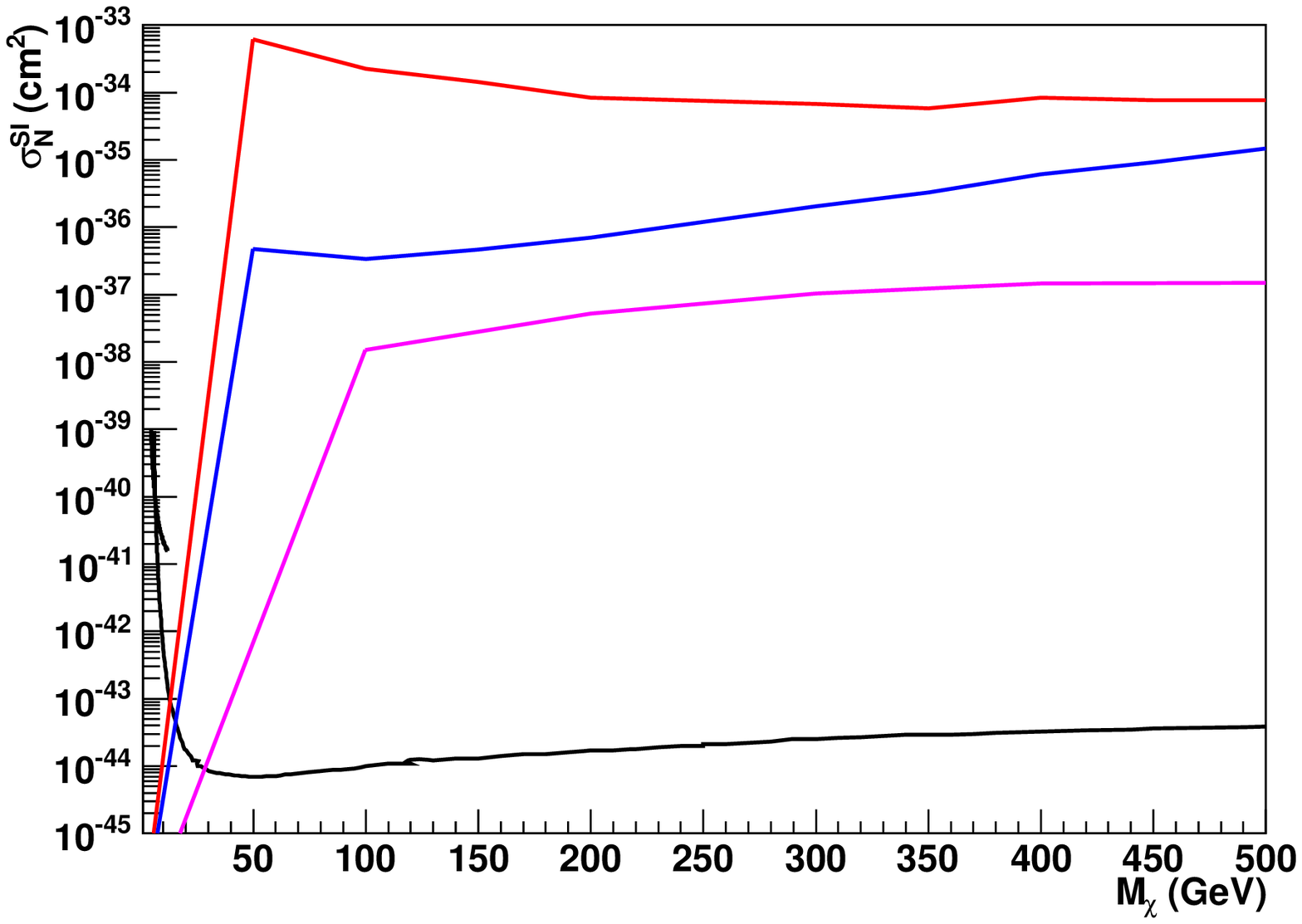}
  \end{center}
  \caption{\label{fig:sigDT1} Constraints on direct detection spin-independent cross section as a function of dark matter mass. The black curves correspond to the experimental limits from Xenon 100 and the CDMSII low-threshold analysis, and the red, blue, and magenta curves correspond to the bounds for operator DT1 mediators of mass $M_\phi$ of $10+M_\chi$, $100+M_\chi$, and 1000 GeV, respectively.}
\end{figure}

\begin{figure}[ht]
  \begin{center}
    \includegraphics[width=.9\textwidth]{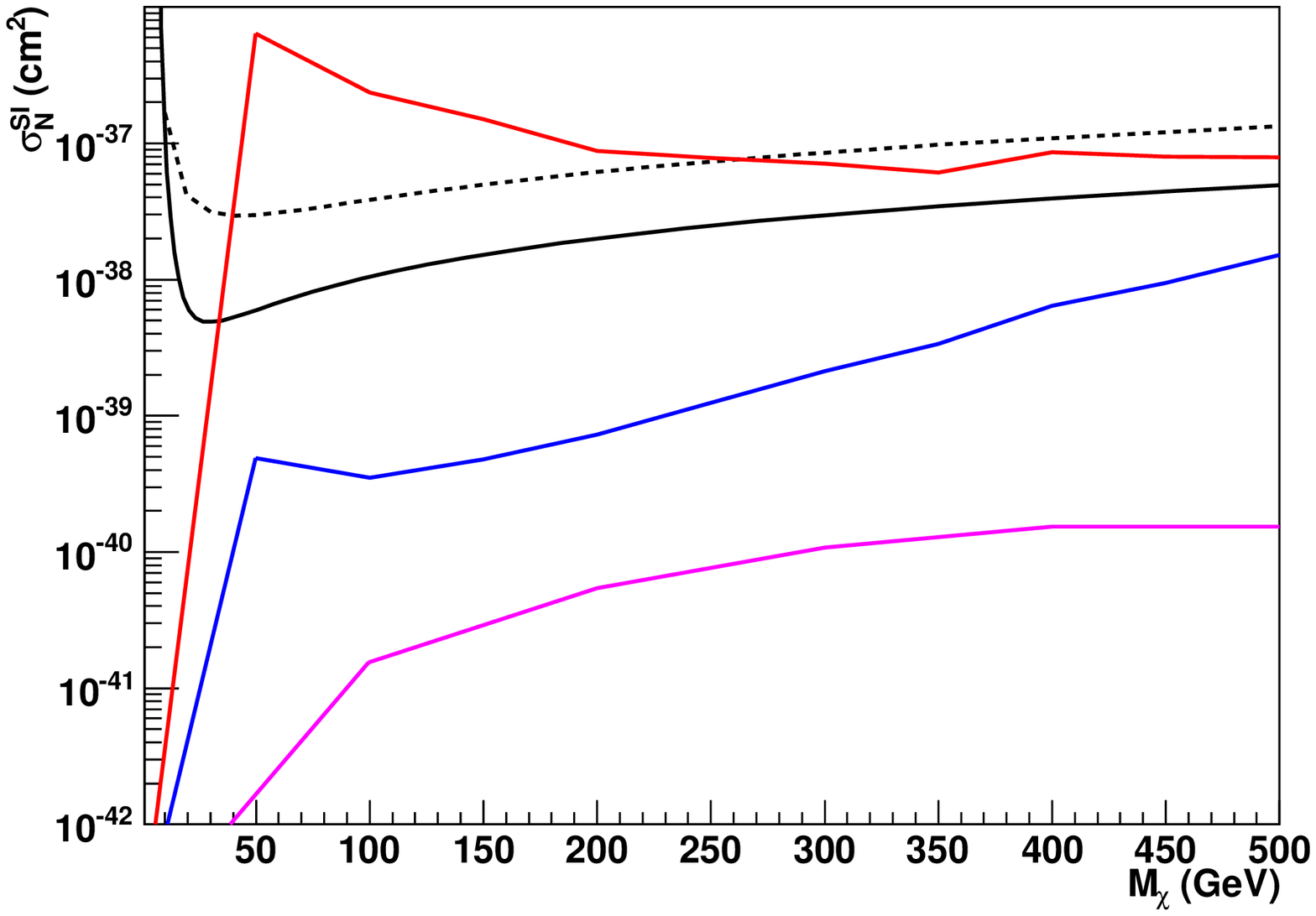}
  \end{center}
  \caption{\label{fig:sigDT1SD} Constraints on direct detection spin-dependent cross section as a function of dark matter mass. The black curves correspond to the experimental limits from Xenon 10 on neutron scatterings (solid curve) and PICASSO on proton scattering (dashed curve), and the red, blue, and magenta curves correspond to the bounds for operator DT1 with mediators of mass $M_\phi$ of $10+M_\chi$, $100+M_\chi$, and 1000 GeV, respectively.}
\end{figure}

\begin{figure}[ht]
  \begin{center}
    \includegraphics[width=.9\textwidth]{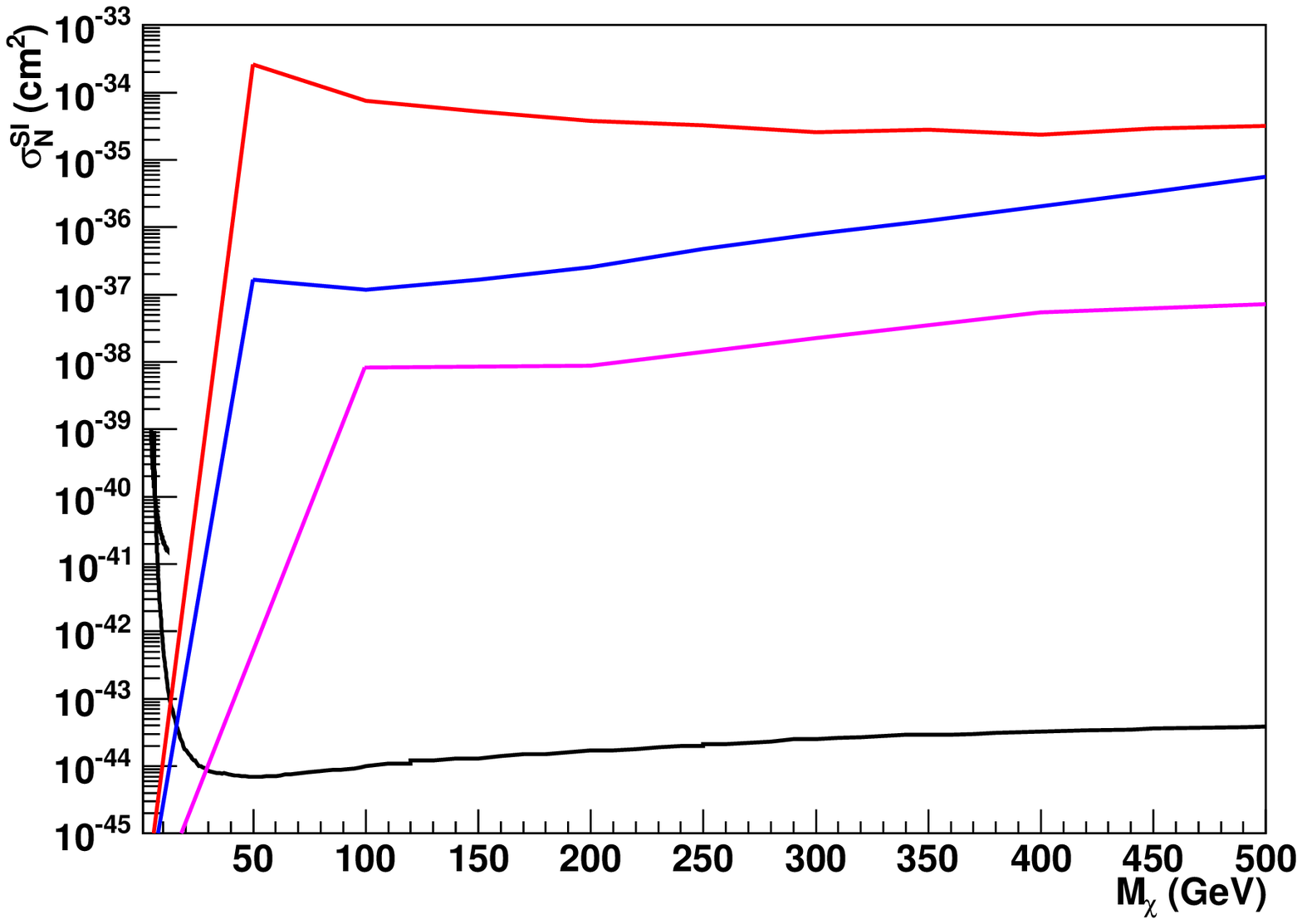}
  \end{center}
  \caption{\label{fig:sigDT2} Constraints on direct detection spin-independent cross section as a function of dark matter mass. The black curves correspond to the experimental limits from Xenon 100 and the CDMSII low-threshold analysis, and the red, blue, and magenta curves correspond to the bounds for operator DT2 with mediators of mass $M_\phi$ of $10+M_\chi$, $100+M_\chi$, and 1000 GeV, respectively.}
\end{figure}

We have translated the bounds on coupling strength to bounds on the direct detection cross section for each model considered.  The results, for various choices of mediator mass, are compared to the most stringent current bounds from direct detection experiments and presented in figures\,\ref{fig:sigDS5}-\ref{fig:sigDT2}.  More detailed spectral 
information from direct detection can also be exploited in an effective theory framework, see \cite{Fan:2010gt}.

\section{Conclusions}
\label{sec:conc}

We have determined the collider bounds due to signals of jets and missing energy on models of dark matter which have UV complete interactions with the SM through mediators within the reach of collider experiments. We have also presented these bounds in terms of the upper limit on direct detection cross sections which they impose for various mediator masses.  We find that having light mediating particles can, in general, degrade the bounds derived based on the assumption of contact interactions, in agreement with previous studies\,\cite{Bai:2010hh,Fortin:2011hv}.


However, the weakening of bounds due to light mediators occurs only within certain kinematic ranges; $M_\phi \lsim 2M_\chi for s-channel and M_\phi \sim M_\chi for t-channel$. Outside of these kinematic ranges we find that bounds are as strong or stronger than those found in previous studies which assumed contact operators. For the purposes of the presented analysis with hard cuts, t-channel mediators must be $\sim$ 300 GeV heavier than dark matter candidates in order to have bounds comparable to those derived from contact operator analyses.

In figures \ref{fig:DS5} and \ref{fig:DS8}, a clear line is visible for $M_\phi \sim 2M_\chi$.  Above this line, resonant enhancement of the dark matter production process can occur. This strengthens the constraints on $M^*$ and therefore the bounds on direct detection cross sections become more stringent. In figures \ref{fig:DT1} and \ref{fig:DT2} a horizontal line is seen, with its position dependent on the hardness of cuts on $\met$   
and jet $P_t$. This is due to the presence or absence of enough mass splitting between the mediator and the dark matter candidate to generate kinematics beyond the cuts without significant initial boost. When the splitting is large compared to the cut values, the limits are largely insensitive to changes in mediator mass.

Though we have only presented limits here for Dirac dark matter, the key differences between contact interaction models as presented in \cite{Goodman:2010ku,Goodman:2010yf,Rajaraman:2011wf,Bai:2010hh,Fox:2011fx}, are not strongly dependent on the Lorentz representation of the dark matter, as it is largely an issue of phase space and kinematics.  These bounds apply to all models of dark matter without regard to the relic density generation mechanism
\cite{Buckley:2011kk}.

\section*{Acknowledgements}
The authors thank Tim Tait and Arvind Rajaraman for helpful discussions.  The work of WS is supported in part by
NSF grant PHY-0970171.  The authors also acknowledge the hospitality of TASI-2011 at the University of Colorado,
where some of this work was undertaken.

\end{document}